\documentclass[12pt]{article}
\usepackage{epsfig}

\begin{document}
\title{Brain neurons as quantum computers: {\it in vivo} support
of background physics}
\author{ A. Bershadskii$^{1}$, E. Dremencov$^2$, J. Bershadskii$^1$
and G. Yadid$^2$}

\maketitle
\begin{center}
{\it $^1$ICAR, P.O. Box 31155, Jerusalem 91000, Israel\\
}
\end{center}
\begin{center}
{\it $^2$Faculty of Life Sciences, Bar-Ilan University,
Ramat-Gan 52900, Israel,}
\end{center}

\begin{abstract}

  The question: whether quantum coherent states can sustain decoherence,
heating and dissipation over time scales comparable to the dynamical
timescales of the brain neurons, is
actively discussed in the last years. Positive answer on this question
is crucial, in particular, for consideration of brain neurons
as quantum computers. This discussion was mainly based on theoretical
arguments. In present paper nonlinear statistical properties
of the Ventral Tegmental Area (VTA) of genetically depressive limbic brain
are studied {\it in vivo} on the Flinders Sensitive Line of rats (FSL).
VTA plays a key role in generation of pleasure and in
development of psychological drug addiction. We found that
the FSL VTA (dopaminergic) neuron signals exhibit multifractal properties for
interspike frequencies on the scales where healthy VTA dopaminergic
neurons exhibit bursting activity.
For high moments the observed multifractal (generalized dimensions) spectrum
coincides with the generalized dimensions spectrum calculated for a spectral
measure of a {\it quantum} system (so-called kicked Harper model,
actively used as a model of quantum chaos). This observation can be considered
as a first experimental ({\it in vivo}) indication in the favour of the quantum
(at least partially) nature of the brain neurons activity.

\end{abstract}

\newpage

\section{Introduction}

  Many authors have argued that consciousness can be understood as
a macroquantum effect. In particular, Penrose \cite{p} proposed
that this takes place in microtubules, the ubiquitous hollow cylinders
that among other things help cells maintain their shapes. It has
been argued that microtubules can process information like a cellular
automaton \cite{h}, and Penrose suggests that they operate as a quantum
computer (see for a review \cite{a}). On the other hand, a number of
other authors have conjectured
that environment-induced decoherence will rapidly destroy the quantum
macrosuperpositions in the brain (see for instance \cite{z},\cite{t} and
references therein). Indeed, the decoherence (as well as dissipation and
heating) is a serious obstacle to all applications exploiting quantum
coherence. Recently, considerable effort has been devoted to designing
strategies able to counteract the undesired effects of the coupling with
an external environment. Notable examples of these strategies in the field
of quantum information are quantum error correction codes,
error avoiding codes \cite{zr}, and "parity kicks" method \cite{vt}.
It is only beginning of such activity. It is quite possible that their exist
still unknown strategies (and processes \cite{mns},\cite{l})
which can eliminate in principle any undesired effect of the environment.
Nature can utilize these strategies (processes) in neurons. Therefore,
the simple calculations of the decoherence characteristic times
(without taking into account such corrective possibilities)
cannot be considered as a final verdict.

 Methods of statistical physics are actively used for investigation of
probabilistic properties of different neurons \cite{ivac}-\cite{b1}.
In the present paper we will study the data obtained
{\it in vivo} for so-called Ventral Tegmental Area (VTA).

  The Ventral Tegmental Area (VTA) is a midbrain nucleus
consisting of the dopaminergic cells.
VTA is known as a part of the limbic brain that corresponds
to such high brain functions as cognition, learning, rewarding
and emotional behavior. VTA plays a key role
in the generation of pleasure and in development of
psychological drug addiction.
This area is also involved in control of the gonadal hormones \cite{cs}.\\

The VTA dopaminergic cells fire in
irregular manner with a mean rare rate between 0.5 to
10 Hz. Firing patterns of the healthy VTA dopaminergic cells
include single spikes and short bursts containing 3-7
spikes. It is believed that the bursting manner of
firing is probably pulled on by rewarding stimuli accepted
from the glutamate neurons originating from prefrontal cortex and
hippocampus. Bursting activity of the normal VTA
dopaminergic cells results in dopamine release in limbic
brain. Dopamine release from VTA dopaminergic cells accompanies
rewarding behavior.
VTA dopaminergic cells are autoregulated via dopamine
autoreceptors expressed on their somas. Additionally,
activity of the VTA dopaminergic cells is regulated by serotonin,
noradrenalin and acetylcholine. \\

  Specific nonlinear
properties of the VTA dopaminergic neurons are actively studied in
recent years (see, for instance \cite{mas},\cite{mas1} and references
therein).
Relationship between these properties and rewarding function of
VTA is one of the topical problems. Therefore, comparative analysis
of the nonlinear properties of the VTA signals generated by normal brains
and ones generated by brains with genetically
suppressed rewarding function of VTA seems to be of significant interest.
For this purpose we use genetically defined rat model of depression
(Flinders Sensitive Rat Line - FSL). Moreover, the genetically defined
depression of the FSL rats (which suppresses the bursting activity of the
dopaminergic neurons) and anaesthesia used in our experiment (see below)
allows us extract a {\it background} signal generated by the dopaminergic
neurons. Using
universality of the background signal we will combine the signals taken from
5 single dopaminergic neurons (3 FSL rats) in order to obtain statistically
representative data set (about 10000 spikes).

  Usually, the time intervals, $T$, between neighboring spikes are used in order
to study statistical properties of neuron signals. In the present paper we
will use {\it inverse} interspike intervals (interspike frequencies)
$\omega = T^{-1}$. While for probability density analysis there is no
advantage in such choice, for a moment (multifractal) analysis the interspike
frequencies turned out to be more informative (see below).
In particular, we will show that the FSL VTA neuron signals exhibit
multifractal properties for
the interspike frequencies on the scales where the healthy VTA
dopaminergic neurons
exhibit bursting activity. The obtained generalized dimensions spectrum
corresponds to the multifractal Bernoulli distribution and for the high
moments this spectrum coincides with generalized dimensions spectrum
calculated for a spectral measure of a {\it quantum} system (so-called
kicked Harper model, actively used as a model of quantum chaos).

\section{Experimental methods and materials}

  Male Sprague-Dawley tars were
used in all experiments. Animals were anaesthetized with
chloral hydrate (400 mg/kg., i.p.) and mounted in
stereotaxic apparatus. The hole was drilled 4.2 mm.
anterior from the interaural line and 1.0 mm lateral from
the medial line. Extracellular recordings were processed by
an electrode from VTA (8.0-8.6 mm dorsal from the lambda).
The constant level of the anaesthesia was checked by EKG
and chloral
hydrate was added  as necessary. Single unit recording
was carried out by amplitude discrimination. Each
recording from the single cell included at least 2000 spike
events. After each experiment, the recording site was
marked by a lesion caused by 15 mA DC for 10 sec.
Brains were removed and stained with formalin before
histological examination. Frozen sections were cut at
50 mm intervals. Microscopic examination of the sections
was carried out aiming to verify that the electrode tip was placed
in VTA.

\section{Multifractal analysis}

 The interspike frequencies: $\omega (n) = T(n)^{-1}$, have obvious
meaning in the case of simple periodic signals with constant period $T$.
For a varying interspike interval $T (n)$ (where $n$ is number of
a spike in the spike series) meaning of $\omega (n)$ is not so clear,
especially for a random-like $T(n)$. Therefore, it is useful to introduce a
measure based on the interspike frequencies through a moving average
$$
\omega_r = \frac{1}{r} \sum_{i=n}^{i=n+r} \omega (i)  \eqno{(1)}
$$
This measure has a simple meaning of an average frequency for
$r$ spikes. Then, one can try to analyze this measure on existence
of scaling properties using its moments:
$$
\langle \omega_r^p \rangle \sim r^{-\mu_p}    \eqno{(2)}
$$
Usually, normalized (dimensionless) moments
$\langle \omega_r^p \rangle/\langle \omega_r \rangle^{p}$
are used for this purpose. In our case, however,
we have $\mu_1=0$ directly from the definition.
Therefore, the scaling exponents of the dimensionless moments
(if existst) coincide with those of the moments (2).
  If the scaling exists one can use the exponents $\mu_p$ in order
to calculate the generalized dimensions $D_p$ \cite{pv}:
$$
D_p=1-\frac{\mu_p}{p-1}        \eqno{(3)}
$$

\begin{figure}[ht]
\epsfig{file=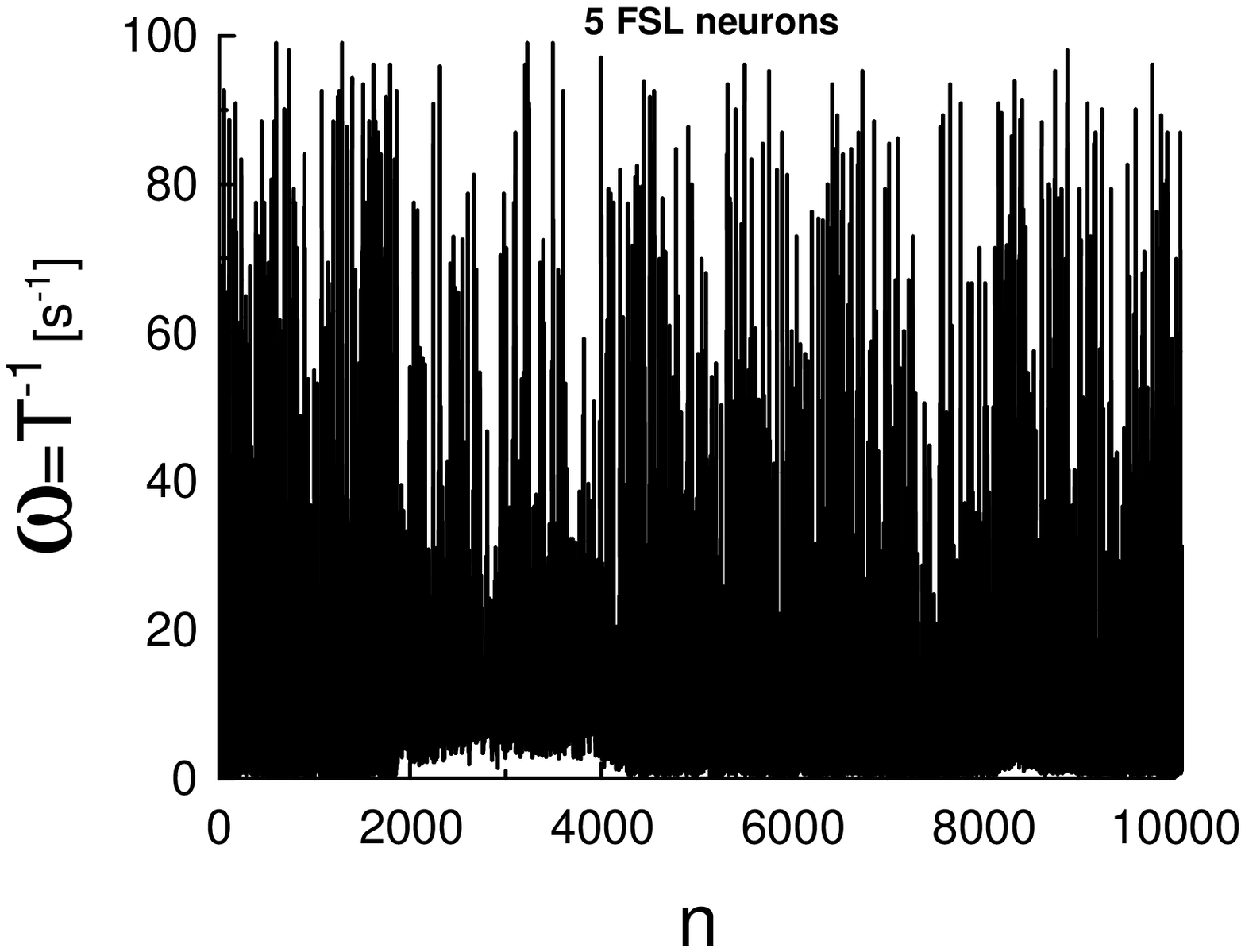,width=4.5in} \caption{\footnotesize
Composite signal, $\omega (n)$, obtained from 5 singular
dopaminergic (VTA) neurons of 3 FSL rats.}
\end{figure}

\begin{figure}[ht]
\epsfig{file=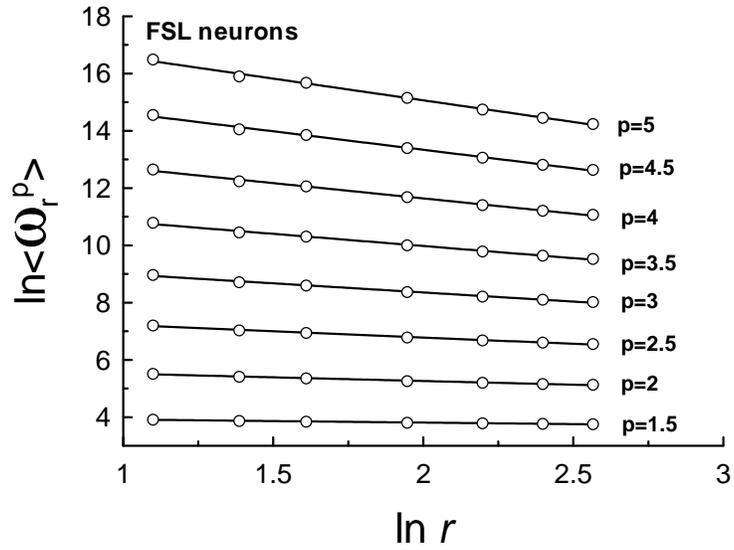,width=4.5in} \caption{\footnotesize
Logarithm of moments (2) versus $\ln r$ for the data set shown in
figure 1, for $r=3-13$. The straight lines (the best fit) are
drawn to indicate scaling (2).}

\end{figure}

\begin{figure}[ht]
\epsfig{file=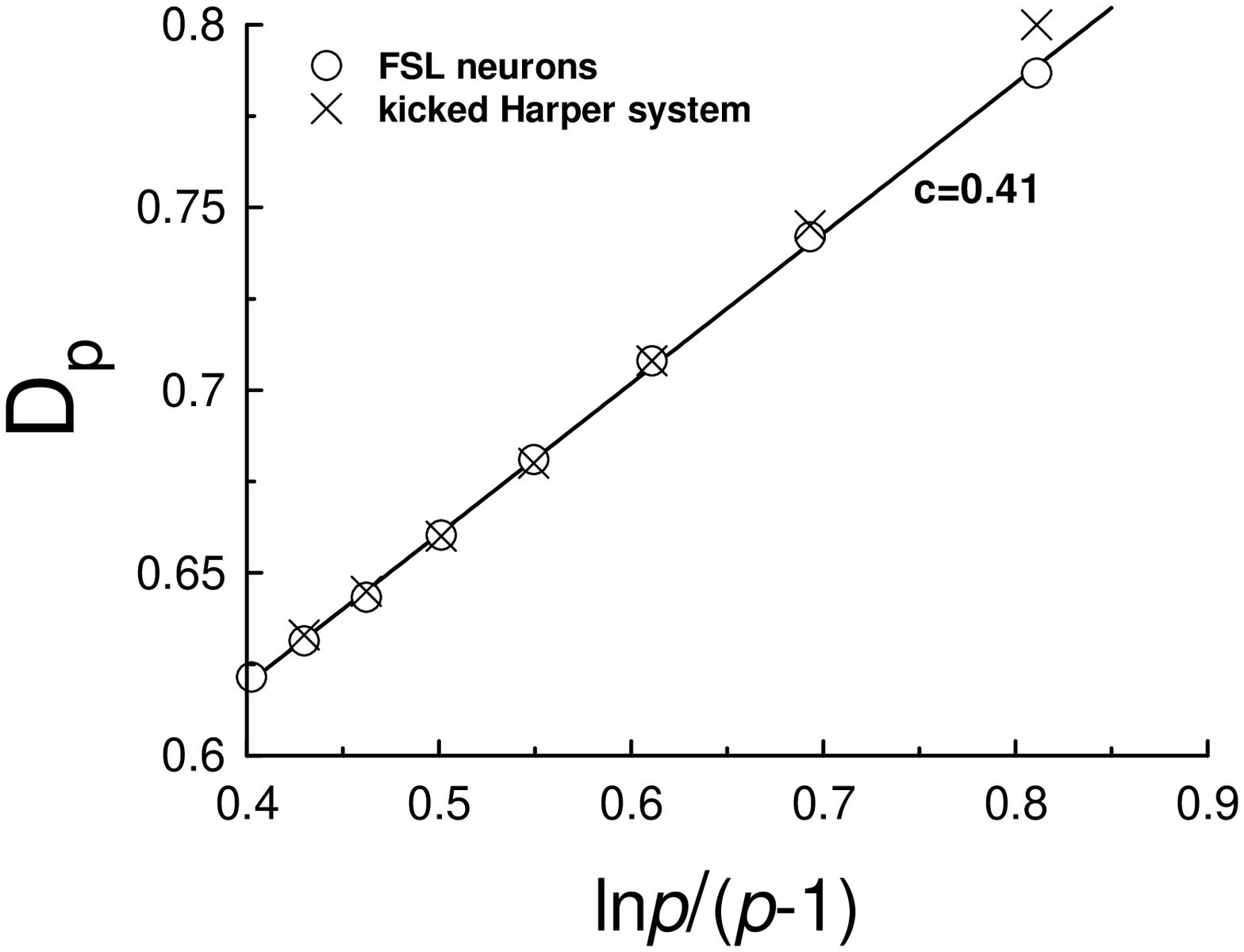,width=4.5in} \caption{\footnotesize
Generalized dimensions $D_p$ versus $\ln p/(p-1)$ calculated using
the exponents $\mu_p$ extracted from figure 2 (open circles) and
for the kicked Harper model \cite{abc} (crosses). The straight
line (the best fit) is drawn in figure 3 to indicate the
multifractal Bernoulli distribution asymptotic (4).}
\end{figure}

  Figure 1 shows composite signal, $\omega (n)$, obtained from 5
singular dopaminergic (VTA) neurons of 3 FSL rats. Because the
genetically defined depression suppresses the bursting activity
of the dopaminergic neurons and because of the anaesthesia used
in our experiment we can expect a {\it background}
signal from the FSL dopaminergic neurons.
In particular, we can expect an universality of the individual
signals (i.e. a statistic similarity even between the signals obtained
from the dopaminergic neurons which belong to different FSL rats).
The combined signal contains of about 10000 data points. A few data
points with relatively short (less than 0.01 s) interspike intervals were
excluded from the combined data set (that is about 1$\%$ of the
entire number of the data points).

   Figure 2 shows the moments (2) calculated for the data set shown
in figure 1, for $r=3-13$. Log-log scales are chosen for comparison with
the scaling equation (2). The straight lines (the best fit) are drawn
to indicate scaling (2) in these scales.

   Figure 3 shows the generalized dimensions $D_p$ (3) (calculated
using the exponents $\mu_p$ extracted from figure 2). The scales:
$D_p$ versus $\ln p/(p-1)$ are chosen in figure 3 for comparison
with the theoretical prediction for the multifractal Bernoulli
distribution \cite{b3}:
$$
D_p = D_{\infty} + c\frac{\ln p}{(p-1)}   \eqno{(4)}
$$
The straight line is drawn in figure 3 to indicate the multifractal
Bernoulli distribution asymptotic (4) with multifractal specific heat
$c \simeq 0.41$.

\section{Discussion}

  The observed scaling interval: $r=3-13$, covers the scales where bursting
activity of the healthy dopaminergic neurons takes place (see Introduction).
Analogous analysis of the control (healthy) rats did not reveal the
multifractal behavior in this interval of scales. This observation
can be interpreted as multifractality of the {\it background} signal of the
dopaminergic (VTA) neurons. Moreover, clear tendency to the multifractal
Bernoulli distribution, seen in figure 3 for the high moments (the straight
line), indicates a fundamental underlying physics (kinetics) \cite{b3}.
In order to proceed further in this direction we also show in figure
3 values of generalized dimensions calculated for a spectral measure of
the kicked Harper model actively used as a model of quantum chaos
\cite{abc}. The kicked Harper model is obtained
upon quantization of the following map:
$$
p_{n+1}=p_n+K \sin (x_n), ~~~~~~~~~~~~x_{n+1}=x_n-L \sin (p_{n+1})
$$
where $K$ and $L$ are some parameters. Canonical quantization thus leads to
the one-period evolution operator
$$
U_{L,K} = \exp [-i\frac{L}{h} \cos (h \nu)] \exp [-i\frac{K}{h} \cos (x)]
$$
where operator $\nu =-i\partial/\partial x$, and $h/2\pi$ has to be
considered as an effective Planck constant, playing a role similar to an
incommensurability parameter in a quasi-periodic system. Quasi-energy
eigenvalues and eigenvectors are determined by
$$
U_{L,K} \psi_{\omega} = \exp [-2\pi i \omega ] \psi_{\omega}
$$
and, if we denote by $\psi_0$ the state in which the system is
prepared at $t=0$, the corresponding spectral measure will be indicated
by $d\eta_{\psi_0} (\omega)$:
its support is contained in the unit interval. The autocorrelation
function is easily expressed in terms of the spectral measure as follows:
$$
C(t) = \langle \psi_0 |U_{L,K}| \psi_0 \rangle = \int_{0}^{1} d\eta_{\psi_0}
(\omega ) \exp [-2\pi i \omega t ]   \eqno{(5)}
$$
  Information on the spectral measure is thus obtained by inversion of (5).

 Multifractal analysis requires a sequence of approximations to the
asymptotic spectral measure $\rho_{i(l)}$, where $\rho_{i(l)}$ is the
probability
attached to the $i$-th ball (of size $l$) covering the support
of the spectral
measure. The generalized dimensions is given by
$$
D_p =\lim_{k \rightarrow \infty} \frac{1}{(p-1)} \frac{\log
\chi_k (p)}{\log l_k}
$$
where
$$
\chi_k (p) = \sum_{j=1}^{N_k} \rho_j(l_k)^p, ~~~~~~~~~~l_k = 1/(2N_k+1)
$$

  Phase diagram of the kicked Harper model is roughly divided into three
regions \cite{abc}, characterized, respectively , by a pure point spectrum
(region A), a purely continuous spectrum (region B), and a mixed spectrum
(region C). Within region A scaling features were consistently not observed.
Region B is, on the other side, characterized by
good scaling properties, and a converging, non-trivial spectrum of
generalized
dimensions were obtained in \cite{abc}. The generalized dimensions for the
parameters $K=2$, $L=5$ (with $N=6400$) are shown
as crosses in figure 3.      \\

  Reported above results can be considered as an indication of existence of
a background dopaminergic neuron signal with universal multifractal
properties for the scales where healthy dopaminergic neurons exhibit
bursting behavior. This signal has multifractal Bernoulli distribution as
high moments asymptotic that may be
related to an underlying physics (kinetics) of mesoscopic systems \cite{b3}.
In particular, the multifractal properties of a spectral measure of the
kicked Harper model are in good quantitative correspondence with the observed
multifractal properties of the dopaminergic neuron signals. The kicked
Harper model is a prominent example of quasi-periodically driven
quantum systems with a chaotic classical analogue, that also can be
of significant interest for the discussion mentioned in Introduction. \\

The authors are grateful to K.R. Sreenivasan for cooperation and to the
Machanaim Center (Jerusalem) for support.

\pagebreak

\end{document}